# BREAST CANCER CLASSIFICATION BASED ON HISTOPATHOLOGICAL IMAGES USING A DEEP LEARNING CAPSULE NETWORK

**Hayder A. Khikani[1], Naira Elazab[2], Ahmed Elgarayhi[1], Mohammed Elmogy[2, *], Mohammed Sallah[1,3]**

[1]Applied Mathematical Physics Research Group, Physics Department, Faculty of Science, Mansoura University, Mansoura 35516, Egypt
[2]Information Technology Department, Faculty of Computers and Information, Mansoura University, Mansoura 35516, Egypt
[3]Higher Institute of Engineering and Technology, New Damietta City, Egypt

**Corresponding author: Mohammed Elmogy (melmogy@mans.edu.eg)**

**The first two authors contributed equally to this work and shared the first authorship.**

**The fourth and fifth authors are sharing the senior authorship.**

## ABSTRACT

Breast cancer is one of the most serious types of cancer that can occur in women. The automatic diagnosis of breast cancer by analyzing histological images (HIs) is important for patients and their prognosis. The classification of HIs provides clinicians with an accurate understanding of diseases and allows them to treat patients more efficiently. Deep learning (DL) approaches have been successfully employed in a variety of fields, particularly medical imaging, due to their capacity to extract features automatically. This study aims to classify different types of breast cancer using HIs. In this research, we present an enhanced capsule network that extracts multi-scale features using the Res2Net block and four additional convolutional layers. Furthermore, the proposed method has fewer parameters due to using small convolutional kernels and the Res2Net block. The suggested model was trained and evaluated using the publicly BreakHis dataset, achieving an accuracy of 95.6% and a recall of 97.2%. As a result, the new method outperforms the old ones since it automatically learns the best possible features. The testing results show that the model outperformed the previous DL methods.

## 1. INTRODUCTION

Breast cancer (BC) predicted has affected 2.3 million women worldwide in 2020, with 685,000 fatalities. Breast cancer had been detected in 7.8 million females in the previous five years as of 2020, making it the most common cancer on the planet. It causes more disability-adjusted life years (DALYs) in women worldwide than any other form of cancer [1]. It strikes women at any age after puberty in every country, with rates rising as they age, according to the world health organization (WHO). It is still a complicated disease, and appropriate systemic grading is important in evaluating a patient who has just been affected by the disease. It consists of five stages. Although the likelihood of survival decreases as the cancer advances to stage IV, these stages are defined by the size of tumors, which are classified as invasive or noninvasive cancer, and have spread to other regions. As a result, early detection and analysis of breast cancer raise survival rates and lower fatality rates [2].

The majority of breast cancers are diagnosed via mammography or ultrasonography; however, they can sometimes be discovered by chance during other tests. On breast computed tomography (CT) images, incidental breast tumors can be discovered. Since chest CT scanning usually includes the breasts, deep learning would increase the sensitivity of unexpected tumor detection on breast CT [3].

Breast examination in these patients is difficult due to significant physiological changes that occur during gestation and breastfeeding. The imaging problems of the breast during gestation and breastfeeding are worth discussing. In each manner and accordance with benign and malignant objects' normal imaging appearance, imaging techniques and imaging properties are presented, along with pathological association and supporting examples [4].

Breast cancer is a serious hazard to women's health. Drug responses related to various breast cancer subtypes show evident implications for therapy outcomes; thus, correct subtype classification is crucial. Breast cancer subtype categorization has lately been investigated using a variety of approaches. Early detection of breast cancer before surgical procedures and correct identification of the subtype can decrease the incidence of surgical treatments and the mortality rate. Limited X-ray mammography seems to be the best model for breast mass monitoring at the moment [5]. The most popular techniques used by specialists to diagnose breast cancer are clinical and pathological diagnostic procedures. Usually, tissue samples acquired from patients using pathological techniques are examined to establish whether tumor formations are present [6].

Because of disruptive technological breakthroughs and outstanding experimental results, widely in the area of image processing and analysis, Artificial intelligence (AI) has lately become a widely popular buzzword. Specialties in medicine that rely heavily on images, such as radiology, pathology, and cancer, have grasped the opportunity and invested heavily in research and development to bring AI's potential to practical use. AI became a more common tool for common medical imaging analytic tasks like diagnosis, segmentation, and classifications [7].

AI techniques in health care have gotten much attention in the previous decade. Computerized medical image processing has been revolutionized. Improvements in machine learning (ML) methodologies, particularly the advent of deep learning (DL) in radiography, have been accompanied by massive gains in computational processing capacity. ML and DL have a lot of potential in image analysis, clinical predictive modeling, and precision medicine [8].

The use of AI for medical pictures enables automated disease diagnosis, histology, stage or subclass characterization, and patient classification based on therapeutic outcome or prognosis. In addition, it allows for highlighting specific regions in images, quantifying component volumes, and extracting characteristics from images. When paired with machine learning techniques, it leads to measuring features or image categorization [9].

Many modeling techniques regarding breast cancer survival prognosis have been presented recently; nevertheless, the majority of them build predictive models based solely on genomic data, with only a few taking into account additional information from pathological imaging [10]. The histological diagnosis of cancer is still regarded as the gold standard despite significant improvements in medical knowledge. As a result, the field requires the development of automated and exact histopathological image analysis systems [11]. In the pathological detection of breast cancer, detecting mitosis in pathological portions is critical because it is used to assess the tumor's competitiveness and provide more thorough and trustworthy information for correct management and therapy. The current popular mitotic detection approaches are categorized into four groups: classic methods, DL techniques, approaches integrating DL and conventional techniques, and additional methods [12].

Mammography, CT, magnetic resonance imaging scans, ultrasonography, and biopsy are some imaging modalities used to gather BC diagnosis samples. Histopathological pictures acquired from biopsies may impact how and when cancer is identified. Pathologists shall use computer-assisted diagnosis technology to aid in the early detection of breast cancer [13].

In this work, we introduced an automated DL technique to classify breast cancer based on histopathological images from 82 women with breast tumors. The DL technique uses the capsule net architecture in which convolutional layers are supplemented with Res2Net blocks, and the capsule layer is based on a linear combination between capsules. The results obtained with the proposed model are better than other recent DL approaches. This demonstrates that our model is a reliable classifier for breast cancer classification based on histological images.

The following points outline the main contributions of our suggested system:
1- To extract multi-scale features and enhance the receptive field of each convolutional layer, we suggest using the Res2Net block according to the model in that research.
2- The four convolutional layers with small kernels are applied to minimize the training parameters.

3- The suggested system is evaluated using a variety of performance criteria. Furthermore, we validated our suggested system by comparing it to some existing systems.
4- Using transfer learning methodologies, the suggested model attained the highest accuracy results.

The rest of the paper will be organized into four sections. Section 2 discusses the current related work. Section 3 presents the proposed method. In Section 4, the experimental results and the discussion are described. Finally, the conclusion is introduced in Section 5.

## 2. RELATED WORK

DL has now surpassed other conventional methods in a wide range of fields. DL approaches have demonstrated outstanding performance in various disciplines, including medical imaging. With the use of image processing methods, it became easier to detect and classify tumors from an infected breast. Liu et al. [14] introduced a DL system to identify breast cancer molecular types. They combined the gene and picture data. Validated repeatedly on a dataset with good accuracy levels, a mixed DL model is developed. A framework for multimodal fusion has been implemented. They produce feature extraction networks to create a multimodal fusion architecture depending on the distinct forms and states. Afterward, they fuse the result of two feature networks using the principle of weighted linear.

Kousalya and Saranya [15] extracted the image features using the convolutional neural network (CNN) DenseNet, which are then supplied into fully connected layers for categorizing cancerous and benign cells. Their models are trained, verified, and evaluated for image categorization of breast cancer. However, Duggento et al. [16] introduced DL methods for cross-domain, cross-discipline diagnosis and classification models that use big and/or complicated real-world datasets. Computational vision problems, particularly image enhancement and interpretation, are where DL architectures shine. This has sparked a flurry of ground-breaking medical imaging applications, and publicly available multicenter image databases have stimulated the growth of DL algorithms tailored to pathology images.

Shi et al. [17] worked on a new unsupervised DL system that has demonstrated its efficacy in feature extraction learning. Conversely, principal component analysis (PCA) is subject to disturbance and outliers, which might degrade PCA network's effectiveness. A required information Grassmann average network (GANet) and quaternion GANet algorithms are being developed to discover effective feature interpretations for histopathology images that include color information.

A new nucleus-guided extraction of features approach has been developed by Zhenga et al. [18] based on CNN. Images are used to detect nuclei, which are subsequently trained using a convolution neural network including three hierarchy structures. These proposed features are tested using a histological image library of breast lesions in a classification experiment. They propose a new nuclei-guided CNN.

Jiang et al. [19] developed a framework to add feature fusion into the architecture for supervised hashing, and then a scalable histopathology image analysis technique is used. It enhances accurate method and classification accuracy significantly. It is also flexible and computationally efficient. Content-based picture retrieval systems have been created as case-based thinking tools to help with this approach.

Vo et al. [20] extracted the most useful sensory information for breast cancer detection using DL models, including convolution layers. These DL methods extract better features than handmade feature extraction methods. It has been demonstrated that those DL models can extract better features than traditional feature extraction methods. They showed applications to tumor histopathology images previously thought difficult to diagnose using traditional methods.

Kumar et al. [21] introduced DL-based techniques for interpreting histopathology pictures of human breast tumors have lately gained favor. They have published a dataset containing canine mammary tumor (CMT) histopathological (CMTHis) scans. They also suggested a visual geometry group (VGG)16-based framework and examined the performance of the hybrid framework with various classifiers, mostly on CMT datasets CMTHis and the breast cancer cell lines dataset BreakHis.

Kurmi et al. [22] enhanced the BC image based on evolutionary histogram equalization and noise removal using Gaussian filtering. The nuclei were enhanced using the proposed ringed kernel, and then nuclei regions were identified using the nuclei attributes and their surroundings. In medium entropy with high complexity (M1) pictures, the suggested technique had trouble segmenting nuclei, resulting in lower accuracy. This technique can segment objects of various sizes using an adaptable kernel size.

Torres et al. [23] provided a processing pipeline for automated segmentation of pigmented BC pictures with various histopathological characteristics. To cope with the enormous size of whole photos, the digital processes were tiled: a substantial portion of the image was divided into patches. After that, each tile was segmented using a deep CNN(DCNN) and an encoder-decoder with a separable aurous convolution structure. Once confirmed, it has proven to be a promising strategy for segmenting problematic image patches.

To test the approach, Iren et al. [24] reported three different histopathology imaging datasets. They claimed that their method is based on pathologists' image classification methodology because it considers the mixture of architectures and textures which can only be understood in the complete image, as opposed to the CNN-based technique's patch-wise classification model. This method has looked at the problem of determining breast cancer's malignancy from the histopathological image analysis perspective.

Wang et al. [25] introduced a DL model that used CNN and capsule network (CapsNet) for BC histopathology image classification based on thorough feature fusion and enhanced routing (FE-BkCapsNet). They started with a unique structure with dual tracks built that can extract convolution characteristics and capsule characteristics simultaneously, as well as integrate somatic and structural features into novel capsules to acquire more discriminative information. Then, by altering the loss function and incorporating the routing paths into the overall optimization technique, routing coefficients are implicitly and adaptively optimized.

Carvalho et al. [26] constructed a model to categorize histopathological breast images into four classes: aggressive carcinoma, carcinoma in situ, healthy tissue, and benign pathology, which employ phylogenetic diversity indexes that define images. One of the most robust classifiers had been used: extreme gradient boosting, random jungle, multilayer perception, plus support vector machine. In addition, they used content-based picture retrieval to verify the classification results and offer a score for unlabeled image sets.

Olaide et al. [27] introduced an architecture that supports knowledgebase formalism and was utilized in a reasoning process for detecting BC. The method allowed the formalization of patient information and domain knowledge using ontologies. The suggested architecture's excellent performance highlights the value of combining guidelines and ontology to represent knowledge. Furthermore, we discovered the applied select and test algorithm's intriguing performance as a hint that it could be used to develop computer-aided diagnostic systems (CADs) for breast cancer detection. Sreeraj and Jesti [28] recommended using nuclear atypia scoring (NAS) to diagnose breast cancer. The suggested cancer detection approach converts every diseased tissue into an artifact. The proposed approach delivers the count of the observed cells and identifies the grade. This aids pathologists in determining whether or not cells are cancerous, as well as the number of each type. On the MITOS-ATYPIA-14 challenge dataset, the proposed model was tested.

Chattopadhyay et al. [29] created dense residual dual-shuffle attention network based on DL. Their model included a channel attention mechanization, which improves the model's capacity to learn complicated visual patterns. This mechanism was inspired by the bottleneck unit of the ShuffleNet design. Despite being trained on a very tiny dataset, the model's tightly connected blocks handle both the overfitting and the vanishing gradient problems. Ahmad et al. [30] used patch selection to accurately categorize breast histopathology pictures using transfer learning on a small set of training photos. In order to extract features, CNN is given patches taken from WSIs. These features are used to pick the discriminative patches, which are then fed into an Efficient-Net architecture that has already been trained on the ImageNet dataset. A support vector machine (SVM) classifier is also trained using features taken from the Efficient-Net architecture. Multiple performance measures show that the suggested model performs better than the standard methods. In [31], a technique for recognizing BC histopathology pictures based on deep semantic features and features from a gray-level co-occurrence matrix (GLCM) is presented. The pre-trained DenseNet201 is used as the fundamental model, and a some of the convolutional layer features from the final dense block is retrieved as deep semantic features. These features are then combined with the three-channel GLCM features, and the SVM is used to classify the data.

There is multiple previous scientific research in the literature on the classification of breast cancer using various models, but they have some drawbacks. Previous research had certain disadvantages, such as a lesser number of images in the dataset, lower accuracy, the need for hand-crafted features, a lack of diverse datasets, no publicly available dataset, and an imbalanced number of images in datasets. We propose an enhanced capsule network to solve the constraints noted above. Some earlier studies used a capsule network, but in this study, we used a customized capsule network model for breast cancer categorization that had not been used in previous studies. The customized model has all the image processing characteristics learned during the training phase. We produced good results by employing a customized CapsNet, as demonstrated in the section of the results in this paper.

## 3. PROPOSED METHOD

The current study aims to classify breast cancer using deep learning so that it can be utilized as an automated tool to aid doctors in their diagnosis. The suggested method feeds histology images into the capsule net architecture provided in that study [32], in which convolutional layers are supplemented with Res2Net blocks. The capsule layer is based on the linear combination method between capsules. In addition, we evaluated the performance of the study on several common convolution-based architectures, such as VGG16 [33], ResNet50 [34], and MobileNet [35], in classifying digital pathology breast images. Figure 1 shows an overview of the suggested model.

- **Preprocessing**

We used H & E staining on the breast histology microscopy in this work. Medical professionals may be better able to see the interior shape of the tissue cells using this universal staining technique. However, ensuring the same dye concentration and amount utilized in each image during the staining procedure is extremely challenging, resulting in variances in the color of the stained histology microscope. Other elements that influence the color of histology photos include the brightness of the ambient light throughout the image acquisition procedure. These color discrepancies in histology images can have a negative impact on CNN training and inference. We must uniform the data in this scenario to reduce color disparities in the histology photos. We use a method in [36] for image processing in this study, which provides an approach for a more generic type of color correction that borrows the color properties of one image from another. This method employs a basic statistical analysis to impose one image's color properties on another, allowing for color correction by selecting an acceptable source image and applying its qualities to another.

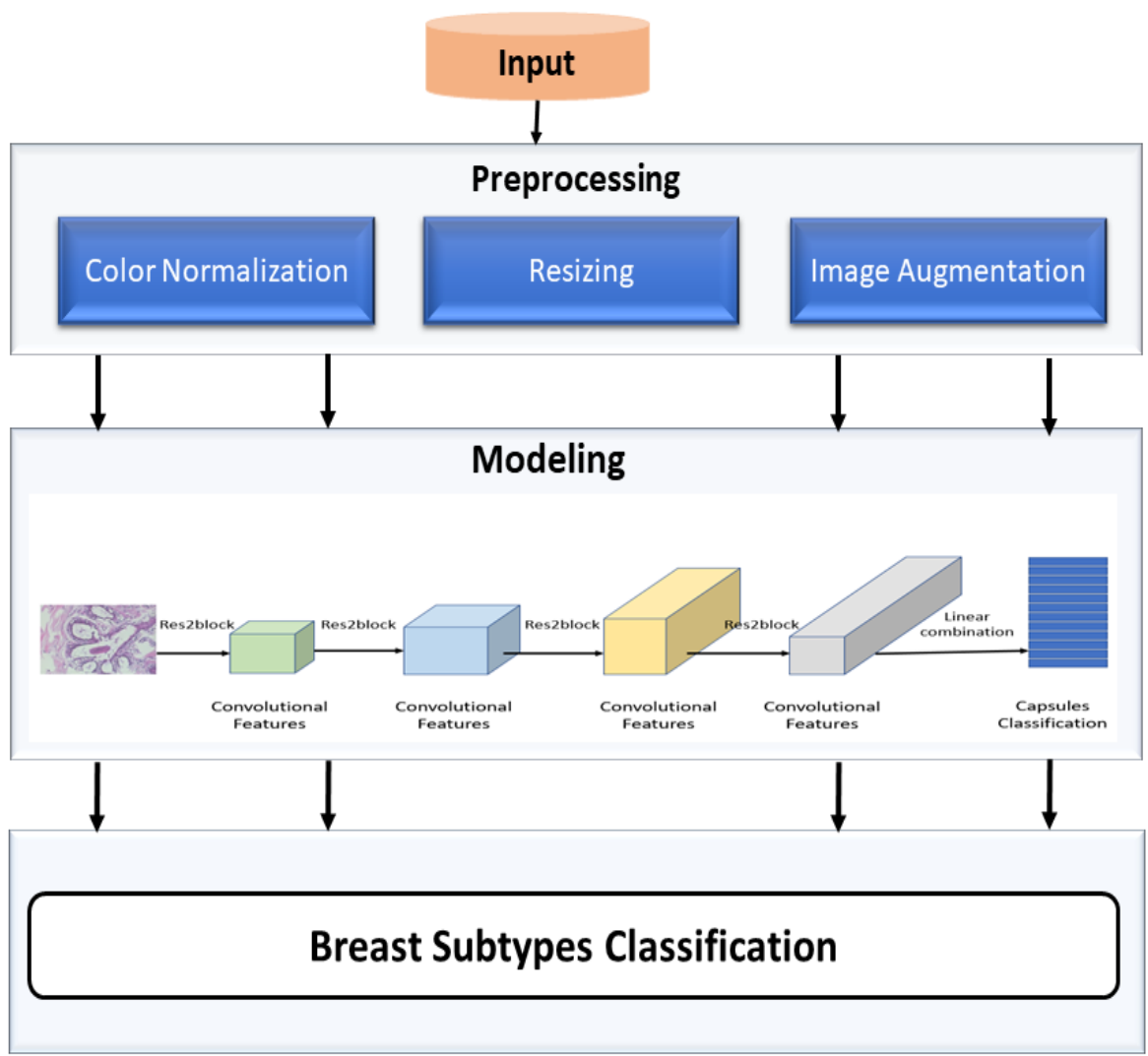

**Figure 1**. The proposed model architecture.

- **Dataset Augmentation**

Larger datasets are required for the training procedure to minimize overfitting. This research used random rotation, flipping, and shearing transformations. Unlike augmentation approaches in [37], the experiment's photos are rotated randomly. In addition, the shearing transformation method is applied, which allows images to be zoomed in or out in multiple directions.

- **Modeling**

We applied the most common DCNN classifiers in image recognition tasks, including VGG16 [33], ResNet50 [34], MobileNetV2 [35], and CapsNet [38], to classify distinct types of breast cancer based on magnification parameters. we compared all the outcomes produced by these models, and the best performing model was chosen as the engine for breast cancer categorization.

**VGG16** In 2014, the visual geometry group at Oxford university entered the ImageNet large-scale visual recognition (ILSVRC) challenge using a DCNN classifier model called VGG [39] and won the classification tasks of the image. Multiple VGG architectures have various levels, two of which are highly common. The first is a 16-layer (VGG16), and the second is a 19-layer (VGG19). In this study, we use VGG16 as a classifier.

**ResNet50** The ResNet architecture was presented in 2015 by He et al. They trained deep convolutional networks using the traditional stochastic gradient descent (SGD) method by residual modules [34]. The ResNet50 model is particularly popular among all types of ResNet models since it has a simpler structure than the other forms, which is why we utilize it in our study.

**MobileNetV2** Another successful DCNN-based classifier technique is Google's MobileNetV2, which was presented in 2018 by Sandler et al. [35]. Because their classifier performs well on benchmarks such as ILSVRC, we chose it as the classifier for this investigation. Although MobileNetV2 is a new notion derived from MobileNetV1 [40], the V2 design has two additional features. The first characteristic is linear bottlenecks between layers, while the second is shortcut connections between bottlenecks.

**Capsule Networks** A capsule is a collection of neurons whose outputs are interpreted as different attributes of the same item. A posture matrix and an activation probability are both included in each capsule. These are activities similar to those of a typical neural network. The likelihood that the capsule entity is present in the current input can be calculated from the length of the output vector of a capsule. The layering of capsules is possible. We employed a layer of cancer capsule (CancerCaps) and a layer of main capsules in our model, which were created by reshaping and compressing the output from the previous convolutional layer. Before the layer of primary capsules, there can be as many convolutional layers as necessary. Only the max-pool layers are absent; instead, convolution with strides greater than 1 was employed to lessen the dimensionality. The output of CancerCaps is utilized to decide the class of the input [38].

- **Advanced Capsule Network**

  A) **Convolutional Layers**

  Only two convolutional layers were applied in the original capsul network [41], but in our enhanced model, the block of Res2Net [42] is applied to improve the convolutional layers' capacity for feature extraction. The four additional convolutional layers are applied to alter the form of the feature maps.

  1) **Block of the Res2net**

  The Res2Net is a variation of the ResNet that employs multiple sets of convolutional operations and creates hierarchical links inside a single block. Unlike layer-wise methods for extracting multi-scale features, the multi-scale features are extracted in the block of Res2Net at the granular level. They can enhance the range of each convolutional layer's receptive fields.

  As illustrated in Ref. [42], the input is initially passed to a set of 1 convolutional kernel, and the resulting feature maps are then split into four groups. Features with varying resolutions are utilized to increase multi-scale capability. To accomplish this, the 3x3 filters of n channels were replaced with a collection of smaller filter groups. These smaller filter groups are linked in a hierarchical style to enhance the number of scales. The input feature maps were divided into several groups. The preceding group of features and another set of input feature maps are transmitted to the next group of filters. This procedure is repeated numerous times until all input feature maps have been processed. Finally, feature maps from all groups are combined and routed to another group of 1 x 1 filters to fuse data.

  2) **Convolutional Architecture**

  We use the Res2Net block because of its good properties. In the enhanced capsule network, there is no skip link between the two adjacent Res2Net blocks; the architecture of the Res2Net block is a component of the convolutional layers. We add four convolutional layers to modify the form of the feature maps. We apply a Res2Net block after each convolutional layer to enhance feature extraction.

  B) **Capsule Layers**

  The primary capsules are immediately utilized to construct the categorization capsules by the original capsule network's principle of using the parts to build the whole. These primary capsules are created from the convolutional feature maps, which only accurately represent the tiniest portions of the identified item and contain many redundant data. We suggest first building the intermediate capsules, which can depict the major components of the identified object, and then combining the intermediate and primary capsules with the final categorization capsules.

## 4. RESULTS AND DISCUSSION

This section describes the experiment's context and the BreakHis database, which is utilized to test the suggested approach.

- **Machine tool**

Keras, Tensor-Flow, and Python toolkits were used to run experiments on the Jupiter notebook. Our system was run on a Dell Workstation 7910 with 2 Intel Xenon E2686v4 18C 36 MB cache. It has 256 GB DDR4 RAM and an NVIDIA RTX 3060 with 12 GB VRAM.

- **Dataset**

A histopathological breast cancer image classification database[43] is used to validate the suggested technique. BreakHis dataset was used in P&D Laboratory in Brazil. It is obtained through an open surgical biopsy, prepared for histological examination, and manually labeled by experienced oncologists. It took 7909 RGB photos in nature size 760 x 460 pixels from 82 ladies suffering from a breast tumor. It includes four magnification factors for each type of benign and malignant tumor. The dataset is divided into 1593 testing photos, 1,263 validation images, and 5,053 training images.

- **Evaluation metrics**

Reliable standards, such as accuracy (ACC), are used to assess and evaluate the proposed system and assure its efficacy. The results were averaged across patch levels rather than patient levels. The confusion matrix parameters are required to determine sensitivity, specificity, and accuracy. In addition to accuracy, we employ sensitivity, specificity, the Dice similarity coefficient (DSC), and the receiver operator characteristic (ROC) curves for performance analysis. The area under the curve (AUC)- ROC curve is obtained by graphing the true positive rate (TPR) versus the false positive rate (FPR) at various threshold points. The true positive rate is also known as recall, sensitivity, or detection probability in machine learning. Similarly, the false positive rate can be determined as specificity. The performance measures used in this paper are provided below.

$$\text{ACC} = \frac{TP+TN}{TP+TN+FP+FN} \tag{1}$$

$$\text{Recall} = \frac{TP}{TP+FN} \tag{2}$$

$$\text{Precision} = \frac{TP}{TP+FP} \tag{3}$$

$$\text{SPE} = \frac{TN}{TN+FP} \tag{4}$$

$$\text{DSC} = 2 * \frac{Precision*Recall}{Precision+Recall} \tag{5}$$

The Matthew correlation coefficient (MCC) calculates the Pearson product-moment correlation coefficient between anticipated and actual values unaffected by the dataset's unbalanced issue.

$$\text{MCC} = \frac{TP*TN-FP*FN}{\sqrt{(TP+FN)(TP+FP)(TN+FN)(TN+FP)}} \tag{6}$$

Certain parameters, such as learning rate, batch size, number of training epochs, and so on, must be carefully selected for the training operation. Choosing the best parameters and values for a deep learning model is crucial. We employ the k-fold cross-validation technique. We train and test the model k times using the training data to get the best representation of the model and parameter values. We use k = 5 in our technique and keep track of the model's accuracy for each fold. The model is trained until it reaches its full potential. The accuracy, precision, recall, and DSC were recorded for each epoch of training and validation. Table 1 depicts the associated values. This table illustrates the model's overall performance on training and validation data because the results in Table1are averaged for each class. Table 1 shows that the model performs remarkably well using ResCapsNet. Aside from accuracy, the precision, recall, and DSC scores are all very encouraging. To validate the efficacy of the suggested model, we compare our results with the results of the state-of-the-art approaches. The suggested model beat the state-of-the-art studies regarding precision, recall, DSC, and accuracy. Figure 2 shows the AUC result of the four DCNN models. In this figure, AUC has been depicted, confirming that ResCapsNet has the highest average AUC compared to the other models.

Table 1. Comparison between different types of CNN models

| CNN Model | ACC (%) | Recall (%) | Precision (%) | DSC (%) | MCC (%) |
|---|---|---|---|---|---|
| VGG16 | 93.1 | 96.02 | 94.2 | 95.5 | 84.8 |
| ResNet50 | 91.3 | 93.20 | 91.4 | 92.3 | 82.3 |
| MobileNet | 92.7 | 95.08 | 92.3 | 93.7 | 83.9 |
| Proposed CapsNet | 95.6 | 97.20 | 94.1 | 93.7 | 90.2 |

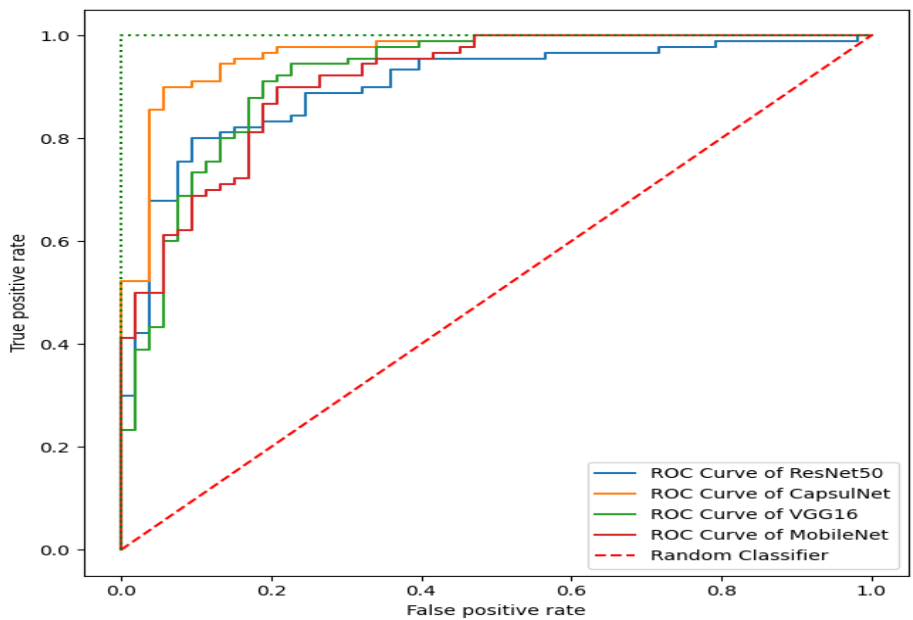

Figure 2. The area under ROC for a VGG16, MobileNet V2, ResNet50, and CapsNet.

▪ **Discussion**

This section details the results of the proposed system in performing the breast cancer classification task in a series of experiments. Confusion matrix analysis was performed to characterize performance further. The confusion matrices of the two-class patch classification of the four magnification variables, accuracy, sensitivity, and specificity, shown in Table 2, demonstrated that our method properly classified the confusion. At the 200x magnification factor, the proposed model attained the best accuracy of 95.6%. Table 3 compares various methodologies from the literature.

The proposed model produced a consistent and accurate recognition rate. The suggested method is stable for multiple Magnification factors, despite the complexity of the classification process due to very fine-grained data. It has the highest accuracy in breast cancer categorization through effort and time, increasing labeling errors. The suggested method for WSI classification significantly benefits retaining all global information about breast cancer images while avoiding the constraints of rectification methods. Although standard correction procedures exist, they have a disadvantage in that the cancer area is just a small percentage of the histological images of breast cancer, resulting in a diagnosis error by the oncologist. As a result, WSI is carefully employed as model input to reduce oncologists' load and increase diagnostic efficiency.

Table 2. The classification performance at different magnifications on the BreakHis dataset.

| Proposed CapsNet | 40x | 100x | 200x | 400x |
|---|---|---|---|---|
| ACC (%) | 93.0 | 91.4 | 95.6 | 89.1 |
| Recall (%) | 92.5 | 90.1 | 97.2 | 87.9 |
| Precision (%) | 91.3 | 91.1 | 94.1 | 87.1 |
| DSC (%) | 91.9 | 90.5 | 93.7 | 87.4 |

Table 3. The comparison of the proposed method with the state-of-the-art methods on the BreakHis dataset.

|  | Method | ACC (%) |
|---|---|---|
| Spanhol et al. [44] | Single task architecture is based on a deep CNN model with a softmax layer. | 85.6 |
| Bayramoglu et al. [45] | A deep CNN model, which is a variant of AlexNet. | 89.7 |
| Kumar et al. [46] | Deep features based on pre-trained AlexNet and VGG16 models. | 94.7 |
| The Proposed Method | Compared to the results of each of the most common DCNN models, the ResCapsNet was chosen for classification. | 95.6 |

## 5. CONCLUSION

Breast cancer is still one of the leading causes of death among women each year.

CAD system has been used to diagnose breast tumors early, quickly, and accurately. The current study is based on breast cancer classification utilizing capsule net architecture. In this paper, a new model built on the capsule network is evaluated. The Res2Net block is used in the proposed capsule network to improve the feature extraction of convolutional layers. The proposed model also has fewer parameters due to using small convolutional kernels and the Res2Net block. On the publicly accessible BreakHis dataset, we conducted a comparative examination of multiple transfer learning models in this study. It has an accuracy of 95.6% in the proposed model. The results demonstrate that this method can be utilized as an automated tool to assist clinicians in disease detection, which may lead to more concentration in therapy in the early stages. For future work, we can combine two or more transfer learning models and extract features from them to achieve better results and develop more robust classifiers with stronger generalization abilities.

## References

[1] Reinert, T., de Souza, A.B., Rosa, M.L., Bedin, S.R., & Barrios, C.H. (2021). Perspectives on the Systemic Staging in Newly Diagnosed Breast Cancer. Clinical breast cancer, 21(4), 309–316.

[2] Michael, E., Ma, H., Li, H., & Qi, S. (2022). An optimized framework for breast cancer classification using machine learning. BioMed Research International, 2022.

[3] Koh, J., Yoon, Y., Kim, S., Han, K., & Kim, E. (2021). Deep Learning for the Detection of Breast Cancers on Chest Computed Tomography. Clinical breast cancer, 22(1), 26–31.

[4] Joshi, S., Dialani, V., Marotti, J.D., Mehta, T., & Slanetz, P.J. (2013). Breast disease in the pregnant and lactating patient: radiological-pathological correlation. Insights into Imaging, 4(5), 527 - 538.

[5] Zhang, L., Li, C., Peng, D., Yi, X., He, S., Liu, F., Zheng, X., Huang, W.E., Zhao, L., & Huang, X. (2021). Raman spectroscopy and machine learning for the classification of breast cancers. Spectrochimica acta. Part A, Molecular and biomolecular spectroscopy, 264, 120300.

[6] Sigirci, I.O., Albayrak, A., & Bilgin, G. (2022). Detection of mitotic cells in breast cancer histopathological images using deep versus handcrafted features. Multim. Tools Appl., 81, 13179-13202.

[7] Barragán-Montero, A.M., Javaid, U., Valdes, G., Nguyen, D., Desbordes, P., Macq, B., Willems, S., Vandewinckele, L., Holmström, M., Löfman, F., Michiels, S., Souris, K., Sterpin, E., & Lee, J.A. (2021). Artificial intelligence and machine learning for medical imaging: A technology review. Physica medica: PM: an international journal devoted to the applications of physics to medicine and biology: official journal of the Italian Association of Biomedical

Physics, 83, 242-256.
[8] Létourneau-Guillon, L., Camirand, D., Guilbert, F., & Forghani, R. (2020). Artificial Intelligence Applications for Workflow, Process Optimization and Predictive Analytics. Neuroimaging clinics of North America, 30 (4), e1-e15.
[9] Avanzo, M., Porzio, M., Lorenzon, L., Milan, L., Sghedoni, R., Russo, G., Massafra, R., Fanizzi, A., Barucci, A., Ardu, V., Branchini, M., Giannelli, M., Gallio, E., Cilla, S., Tangaro, S.S., Lombardi, A., Pirrone, G., de Martin, E., Giuliano, A., Belmonte, G., Russo, S., Rampado, O., & Mettivier, G. (2021). Artificial intelligence applications in medical imaging: A review of the medical physics research in Italy. Physica Medica: PM: an international journal devoted to the applications of physics to medicine and biology : official journal of the Italian Association of Biomedical Physics, 83, 221-241 .
[10] Sun, D., Li, A., Tang, B., & Wang, M. (2018). Integrating genomic data and pathological images to effectively predict breast cancer clinical outcome. Computer methods and programs in biomedicine, 161, 45-53.
[11] Yan, R., Ren, F., Wang, Z., Wang, L., Zhang, T., Liu, Y., Rao, X., Zheng, C., & Zhang, F. (2019). Breast cancer histopathological image classification using a hybrid deep neural network. Methods, 173, 52–60.
[12] Pan, X., Lu, Y., Lan, R., Liu, Z., Qin, Z., Wang, H., & Liu, Z. (2021). Mitosis detection techniques in H&E stained breast cancer pathological images: A comprehensive review. Comput. Electr. Eng., 91, 107038.
[13] Kaushal, C., Bhat, S., Koundal, D., & Singla, A. (2019). Recent Trends in Computer Assisted Diagnosis (CAD) System for Breast Cancer Diagnosis Using Histopathological Images. IRBM, 40 (4), 211–227.
[14] Liu, T., Huang, J., Liao, T., Pu, R., Liu, S., & Peng, Y. (2021). A Hybrid Deep Learning Model for Predicting Molecular Subtypes of Human Breast Cancer Using Multimodal Data. Irbm, 1, 1–13.
[15] Kousalya, K., & Saranya, T. (2021). Improved the detection and classification of breast cancer using hyper parameter tuning. Materials Today: Proceedings.
[16] Duggento, A., Conti, A., Mauriello, A., Guerrisi, M., & Toschi, N. (2020). Deep computational pathology in breast cancer. Seminars in cancer biology, 72, 226–237.
[17] Shi, J., Zheng, X., Wu, J., Gong, B., Zhang, Q., & Ying, S. (2019). Quaternion Grassmann average network for learning representation of histopathological image. Pattern Recognit., 89, 67-76.
[18] Zheng, Y., Jiang, Z., Xie, F., Zhang, H., Ma, Y., Shi, H., & Zhao, Y. (2017). Feature extraction from histopathological images based on nucleus-guided convolutional neural network for breast lesion classification. Pattern Recognit., 71, 14-25.
[19] Jiang, M., Zhang, S., Huang, J., Yang, L., & Metaxas, D.N. (2016). Scalable histopathological image analysis via supervised hashing with multiple features. Medical image analysis, 34, 3-12.
[20] Vo, D.M., Nguyen, N., & Lee, S. (2019). Classification of breast cancer

histology images using incremental boosting convolution networks. Inf. Sci., 482, 123-138.
[21] Kumar, A., Singh, S.K., Saxena, S., Lakshmanan, K., Sangaiah, A.K., Chauhan, H., Shrivastava, S., & Singh, R.K. (2020). Deep feature learning for histopathological image classification of canine mammary tumors and human breast cancer. Inf. Sci., 508, 405-421.
[22] Kurmi, Y., Chaurasia, V., & Kapoor, N. (2020). Design of a histopathology image segmentation algorithm for CAD of cancer. Optik, 218, 164636.
[23] Torres, B.M., Morillo, D.S., Granero, M.Á., & García-Rojo, M. (2020). Automatic segmentation of whole-slide H&E stained breast histopathology images using a deep convolutional neural network architecture. Expert Syst. Appl., 151, 113387.
[24] Fondón, I., Sarmiento, A., García, A.I., Silvestre, M., Eloy, C., Polónia, A., & Aguiar, P. (2018). Automatic classification of tissue malignancy for breast carcinoma diagnosis. Computers in biology and medicine, 96, 41-51.
[25] Wang, P., Wang, J., Li, Y., Li, P., Li, L., & Jiang, M. (2021). Automatic classification of breast cancer histopathological images based on deep feature fusion and enhanced routing. Biomed. Signal Process. Control., 65, 102341.
[26] Carvalho, E.D., Filho, A.O., Silva, R.R., Araújo, F.H., Diniz, J.O., Silva, A.C., Paiva, A.C., & Gattass, M. (2020). Breast cancer diagnosis from histopathological images using textural features and CBIR. Artificial intelligence in medicine, 105, 101845.
[27] Oyelade, O.N., Aghiomesi, I.E., Najeem, O., & Sambo, A.A. (2021). A Semantic Web Rule and Ontologies Based Architecture for Diagnosing Breast Cancer Using Select and Test Algorithm. Computer Methods and Programs in Biomedicine Update, 1, 100034.
[28] Madhavankutty, S., & Joy, J. (2021). A machine learning based framework for assisting pathologists in grading and counting of breast cancer cells. ICT Express, 7, 440-444.
[29] Chattopadhyay, S., Dey, A., Singh, P.K., & Sarkar, R. (2022). DRDA-Net: Dense residual dual-shuffle attention network for breast cancer classification using histopathological images. Computers in biology and medicine, 145, 105437.
[30] Ahmad, N., Asghar, S., & Gillani, S.A. (2022). Transfer learning-assisted multi-resolution breast cancer histopathological images classification. *Vis. Comput., 38*, 2751-2770.
[31] Hao, Y., Zhang, L., Qiao, S., Bai, Y., Cheng, R., Xue, H., Hou, Y., Zhang, W., & Zhang, G. (2022). Breast cancer histopathological images classification based on deep semantic features and gray level co-occurrence matrix. PLoS ONE, 17.
[32] Yang, S., Lee, F., Miao, R., Cai, J., Chen, L., Yao, W., Kotani, K., & Chen, Q. (2020). RS-CapsNet: An Advanced Capsule Network. IEEE Access, 8, 85007-85018.
[33] Simonyan, K., & Zisserman, A. (2015). Very Deep Convolutional Networks

for Large-Scale Image Recognition. CoRR, abs/1409.1556.
[34] He, K., Zhang, X., Ren, S., & Sun, J. (2016). Deep Residual Learning for Image Recognition. 2016 IEEE Conference on Computer Vision and Pattern Recognition (CVPR), 770-778.
[35] Sandler, M., Howard, A.G., Zhu, M., Zhmoginov, A., & Chen, L. (2018). MobileNetV2: Inverted Residuals and Linear Bottlenecks. 2018 IEEE/CVF Conference on Computer Vision and Pattern Recognition, 4510-4520.
[36] Cahill, L.C., Giacomelli, M.G., Yoshitake, T., Vardeh, H.G., Faulkner-Jones, B.E., Connolly, J.L., Sun, C., & Fujimoto, J.G. (2018). Rapid virtual H&E histology of breast tissue specimens using a compact fluorescence nonlinear microscope. Laboratory investigation; a journal of technical methods and pathology, 98, 150 - 160.
[37] van Dyk, D. A., & Meng, X.-L. (2001). The Art of Data Augmentation. Journal of Computational and Graphical Statistics, 10, 1–50.
[38] Rajasegaran, J., Jayasundara, V., Jayasekara, S., Jayasekara, H., Seneviratne, S., & Rodrigo, R. (2019). DeepCaps: Going Deeper with Capsule Networks. 2019 IEEE/CVF Conference on Computer Vision and Pattern Recognition (CVPR), 10717–10725.
[39] Russakovsky, O., Deng, J., Su, H., Krause, J., Satheesh, S., Ma, S., … Fei-Fei, L. (2015). ImageNet Large Scale Visual Recognition Challenge. International Journal of Computer Vision, 115, 211–252.
[40] Howard, A. G., Zhu, M., Chen, B., Kalenichenko, D., Wang, W., Weyand, T., Adam, H. (2017). MobileNets: Efficient Convolutional Neural Networks for Mobile Vision Applications. ArXiv, abs/1704.04861.
[41] Sabour, S., Frosst, N., & Hinton, G. E. (2017). Dynamic Routing Between Capsules. ArXiv, abs/1710.09829.
[42] Gao, S., Cheng, M.-M., Zhao, K., Zhang, X., Yang, M.-H., & Torr, P. H. S. (2021). Res2Net: A New Multi-Scale Backbone Architecture. IEEE Transactions on Pattern Analysis and Machine Intelligence, 43, 652–662.
[43] Spanhol, F. A., Oliveira, L., Petitjean, C., & Heutte, L. (2016). A Dataset for Breast Cancer Histopathological Image Classification. IEEE Transactions on Biomedical Engineering, 63, 1455–1462.
[44] Spanhol, F.A., Oliveira, L., Petitjean, C., & Heutte, L. (2016). Breast cancer histopathological image classification using Convolutional Neural Networks. 2016 International Joint Conference on Neural Networks (IJCNN), 2560-2567.
[45] Bayramoglu, N., Kannala, J., & Heikkilä, J. (2016). Deep learning for magnification independent breast cancer histopathology image classification. 2016 23rd International Conference on Pattern Recognition (ICPR), 2440–2445.
[46] Kumar, M.D., Babaie, M., Zhu, S., Kalra, S., & Tizhoosh, H.R. (2017). A comparative study of CNN, BoVW and LBP for classification of histopathological images. 2017 IEEE Symposium Series on Computational Intelligence (SSCI), 1-7.